\documentclass[aps,prc,twocolumn,superscriptaddress,floatfix,nofootinbib]{revtex4-2}

\usepackage{graphicx}
\usepackage{amsmath}
\usepackage{amssymb}
\usepackage[hypertexnames=false]{hyperref}
\usepackage{xcolor}
\usepackage{multirow}
\usepackage{float}

\newcommand{\sqrts}{\sqrt{s_{\rm NN}}}
\newcommand{\Tkin}{T_{\rm kin}}

\newcommand{\bs}{\beta_s}
\newcommand{\betaavg}{\langle\beta\rangle}
\newcommand{\pT}{p_{\rm T}}
\newcommand{\mT}{m_{\rm T}}
\newcommand{\dchi}{\Delta\chi}%
\newcommand{\Tstrut}{\rule{0pt}{2.6ex}}
\newcommand{\Bstrut}{\rule[-1.2ex]{0pt}{0pt}}

\begin{document}

\title{Kinetic freeze-out of the $\phi(1020)$ meson in Pb--Pb collisions
at $\sqrts = 2.76$~TeV: a blast-wave and hybrid-model study}

\author{Neeraj}
\email{neeraj.neeraj@cern.ch}
\author{Amal Sarkar}
\email{amal.sarkar@cern.ch}
\affiliation{Indian Institute of Technology Mandi, Kamand,
Himachal Pradesh, India}

\date{\today}

\begin{abstract}
Hadrons with small hadronic cross sections are expected to decouple from the
expanding fireball earlier than the bulk. We test this for the $\phi(1020)$
meson and the multistrange baryons in Pb--Pb collisions at
$\sqrts = 2.76$~TeV, applying the same blast-wave analysis to published
ALICE spectra of all species. In 0--5\% central collisions the $\phi$
prefers a smaller mean transverse velocity than pions, kaons and protons,
$\betaavg_\phi = 0.602^{+0.036}_{-0.072}$ against $0.651 \pm 0.004$. A
nested likelihood-ratio test of common freeze-out gives $2.1\sigma$
($0.9$--$3.0\sigma$ under fit-range and uncertainty variations). The
tension is driven by the protons and is absent in 10--30\% central
collisions. A vHLLE+SMASH hybrid calculation analyzed in the same way
predicts a $\phi$ flow velocity 0.034 below that of the bulk, from early
decoupling, soft regeneration and in-medium decay losses; soft
regeneration alone could account for most of the observed tension. In
0--5\%, 0--10\% and 20--40\% central collisions the data show half to
two-thirds of the predicted separation; under the model, a separation this
small occurs with probability $P = 0.006$--$0.04$.
The $\Xi$ and $\Omega$, which SMASH does not regenerate, are separated
from the $\pi/K/p$ freeze-out at high significance ($\dchi^2 \geq 51$ for
the $\Omega$), again anchored by the protons. The test passes in the
decays-only limit of the model, and the full calculation reproduces the
size of the separation in 0--10\% and 20--40\%. The multistrange baryons thus show
species-dependent kinetic decoupling; for the $\phi$ the data constrain,
but do not establish, differential freeze-out.
\end{abstract}

\maketitle

\section{Introduction}
\label{sec:intro}

The hadronic stage of an ultrarelativistic heavy-ion collision begins at
chemical freeze-out, close to the pseudocritical temperature
$T_{pc} \approx 156$~MeV~\cite{Borsanyi:2020,HotQCD:2019,Andronic:2018}, and
ends when elastic and pseudoelastic rescattering stops~\cite{Heinz:2007in}. The kinetic
freeze-out temperature $\Tkin$ and the mean transverse expansion velocity
$\betaavg$ are commonly extracted from simultaneous blast-wave fits to the
$\pi$, $K$ and $p$ spectra~\cite{Schnedermann:1993ws,Retiere:2003kf,ALICE:2013mez,STAR:2008med,STAR:2017sal}.
In central Pb--Pb collisions at $\sqrts = 2.76$~TeV, ALICE reports
$\Tkin = 95 \pm 10$~MeV and $\betaavg = 0.651 \pm 0.020$~\cite{ALICE:2013mez}.

Whether all hadrons decouple from the same freeze-out hypersurface is a
long-standing question. Hadrons with small cross sections in a pion- and
nucleon-dominated medium may stop interacting earlier and keep a higher
effective temperature or a smaller flow velocity. Transport calculations
predicted this ``sequential'' freeze-out for multistrange
baryons~\cite{vanHecke:1998yu,Nonaka:2007,Xie:2009ut,Takeuchi:2015ana}.
Slope systematics at the SPS and RHIC have been interpreted in this
way~\cite{NA57:2006jmp,Xu:2001zj,STAR:2003jis}, and coalescence
analyses of multistrange elliptic flow assume a decoupling close to the
hadronization transition~\cite{He:2010vw}.
The $\phi(1020)$ is a particularly clean test case. Its mass is close to the
proton mass, so mass-ordering effects of collective flow are almost the same
for both particles, while its OZI-suppressed coupling to non-strange hadrons
implies a much smaller hadronic cross section~\cite{Shor:1984ui}. For
this reason $\phi$ production and elliptic flow at RHIC were taken as
evidence of the coalescence of thermalized strange quarks and of partonic
collectivity~\cite{Adams:2004ep_phi,STAR:2007mum}. Hybrid calculations show that
the $\phi$ does not fully take part in the radial flow built up in the
hadronic stage, which breaks the mass ordering of elliptic
flow~\cite{Hirano:2007ei}, and have been used to predict the $\phi$ spectra
at the LHC~\cite{Song:2013qma}.

At the LHC, Melo and Tom\'a\v{s}ik reconstructed the final state of central
Pb--Pb collisions with blast-wave fits to many species and found indications
that the $\phi$ and multistrange baryons decouple earlier than the
bulk~\cite{Melo:2015bwa}. Tsallis fits to the same spectra indicate a
kinetic freeze-out temperature that rises with the hadron
mass~\cite{Lao:2016epa}, and the $\Omega$ and $\phi$ spectra obey a
constituent-quark-number scaling that reflects the strange-quark
distribution at hadronization~\cite{Song:2019ncm}. ALICE, on the other hand,
observed that the $p/\phi$ ratio in central collisions is nearly independent
of $\pT$ up to $\sim 3$~GeV/$c$, as expected if the shapes of the two spectra
are governed by common collective flow~\cite{ALICE:2014jbq_phi}. These two
observations pull in opposite directions, and their quantitative
compatibility has not been tested. The hadronic phase is long enough to
matter: pion femtoscopy gives a decoupling time of about
10~fm/$c$~\cite{ALICE:2011dyt}, and the suppression of the $K^{*0}/K$
ratio in central collisions, absent for the longer-lived $\phi$, is
attributed to the rescattering of resonance decay
products~\cite{ALICE:2014jbq_phi,ALICE:2019xyr,Bleicher:2002dm,Knospe:2015nva}.

In this paper we pose the question as a hypothesis test with explicit
treatment of the experimental uncertainties. We (i) fit the bulk and the
$\phi$ with one blast-wave implementation, matched to the published
spectrum convention of each species and averaged over the published
$\pT$ bins;
(ii) test a common freeze-out with a nested likelihood-ratio test in which
the bulk parameters are refitted together with the $\phi$; (iii) propagate
the bulk fit-range and species variations used by ALICE and several
correlation models for the $\phi$ systematic uncertainties; (iv) check
consistency with the measured $p/\phi$ ratio and with $\phi$ regeneration;
(v) repeat the analysis in nine centrality classes; and (vi) confront the
result with a viscous-hydrodynamics plus hadronic-transport
(vHLLE+SMASH) calculation analyzed in exactly the same way, and use a
SMASH box calculation to separate the effects of elastic rescattering,
regeneration, and the selection of a $\phi$ population on its apparent
temperature; and (vii) apply the same test to the $\Xi$ and $\Omega$,
which are not regenerated in SMASH, as a reference for the $\phi$.

The paper is organized as follows. Section~\ref{sec:method} describes the
data, the blast-wave model and the statistical tests.
Section~\ref{sec:bulk} presents the bulk reference fit and
Sec.~\ref{sec:phi} the central-collision $\phi$ analysis and its systematic
variations. Section~\ref{sec:cent} reports the centrality dependence.
Section~\ref{sec:transport} contains the transport-model studies and
Sec.~\ref{sec:multistrange} the multistrange baryons, and
Sec.~\ref{sec:discussion} discusses the results.

\section{Data and method}
\label{sec:method}

\subsection{Data}

We use the ALICE $\pi^\pm$, $K^\pm$, $p$ and $\bar p$ invariant yields in
Pb--Pb collisions at $\sqrts = 2.76$~TeV~\cite{ALICE:2013mez} and the
$\phi(1020)$ $\pT$ distributions $d^2N/(d\pT\,dy)$~\cite{ALICE:2014jbq_phi},
both at midrapidity ($|y| < 0.5$) and taken from HEPData. The two
measurements use identical centrality classes (0--5\%, 5--10\%,
10--20\%, \ldots, 70--80\%). The bulk spectra are corrected for
weak-decay feed-down~\cite{ALICE:2013mez}. For each bulk data point we use
the statistical and systematic uncertainties added in quadrature, as in the
ALICE blast-wave analysis.

The $\phi$ uncertainties are published in four components: statistical,
$\pT$-uncorrelated systematic, a 0.5\% normalization uncertainty, and a
10.05\% ``other $\pT$-correlated'' systematic uncertainty. The last
component dominates. Its bin-to-bin correlation is not published, and it
turns out to control the significance of the result. We therefore treat it
explicitly with several correlation models (Sec.~\ref{sec:cov}). The
nominal $\phi$ fit window contains the six bins with centers
$0.9$--$3.5$~GeV/$c$ (bin edges 0.8--4.0~GeV/$c$). The two
remaining published bins (0.5--0.8 and 4--5~GeV/$c$) enter the
fit-window variations.

\subsection{Blast-wave model}

The Boltzmann--Gibbs blast-wave (BGBW)
yield~\cite{Schnedermann:1993ws,Florkowski:2006} is
\begin{align}
\frac{d^2N}{2\pi\pT\,d\pT\,dy} = A \int_0^1 \xi\,d\xi\;\mT\,
&I_0\!\left(\frac{\pT\sinh\rho}{T}\right) \notag\\
\times\, &K_1\!\left(\frac{\mT\cosh\rho}{T}\right),
\label{eq:bgbw}
\end{align}
with $\mT = (\pT^2 + m^2)^{1/2}$, $\rho(\xi) = \tanh^{-1}(\bs\xi^n)$ and
$\betaavg = 2\bs/(2+n)$. The radial integral is evaluated with 48-point
Gauss--Legendre quadrature. Two details matter for the $\phi$:
\begin{itemize}
\item \emph{Jacobian.} The $\phi$ data are $d^2N/(d\pT\,dy)$, so the model
is multiplied by $2\pi\pT$ before comparison. Omitting this factor changes
the fitted $\phi$ parameters substantially (Appendix~\ref{app:validation}).
\item \emph{Bin averaging.} The $\phi$ bins are up to 1~GeV/$c$ wide. The
model is averaged over each bin (12-point Gauss--Legendre in $\pT$), not
evaluated at the bin center. The bulk spectra are treated the same way.
\end{itemize}
The normalization $A$ of each spectrum is left free; for given shape
parameters its best value is obtained directly from a linear fit.

\subsection{Covariance of the \texorpdfstring{$\phi$}{phi} spectrum}
\label{sec:cov}

Let $\sigma^{\rm u}_i$ denote the statistical and $\pT$-uncorrelated
systematic uncertainties in quadrature, and let $s_i$ and $\nu_i$ denote the
$\pT$-correlated and normalization components. We consider
\begin{align}
\text{quadrature:}\quad & C_{ij} = \delta_{ij}\,[(\sigma^{\rm u}_i)^2 + s_i^2 + \nu_i^2], \notag\\
\text{smooth}(L):\quad & C_{ij} = \delta_{ij}(\sigma^{\rm u}_i)^2
 + s_i s_j\, e^{-(p_i - p_j)^2/2L^2} + \nu_i\nu_j, \notag\\
\text{tilt:}\quad & C_{ij} = \delta_{ij}(\sigma^{\rm u}_i)^2 + u_i s_i\,u_j s_j + \nu_i\nu_j, \notag\\
\text{correlated:}\quad & C_{ij} = \delta_{ij}(\sigma^{\rm u}_i)^2 + s_i s_j + \nu_i\nu_j,
\end{align}
where $p_i$ is the bin center and $u_i \in [-1, 1]$ varies linearly across
the fit window. The correlated case is a pure scale uncertainty. It is
largely absorbed by the free normalization and is therefore the
\emph{least} conservative choice for a shape test. The tilt and
finite-$L$ cases allow the systematic uncertainty to change the spectral
slope. We adopt the quadrature treatment as the baseline and report the
others as a systematic variation of the test itself.

\subsection{Statistical tests}

\emph{Fixed-point test.} The $\phi$ is fitted with free $(T, \betaavg, A)$ at
the bulk value of $n$, and
$\dchi^2_{\rm fp} = \chi^2_\phi(T_{\rm bulk}, \betaavg_{\rm bulk}) - \chi^2_{\phi,\rm min}$
is evaluated at the bulk best-fit point. This test ignores the uncertainty
of the bulk point.

\emph{Joint nested test.} The bulk spectra and the $\phi$ are fitted
simultaneously under two hypotheses. Under $H_0$ all species share
$(T, \betaavg, n)$. Under $H_1$ the $\phi$ has its own $(T_\phi, \betaavg_\phi)$
while $n$ is common; a different flow profile for the $\phi$, which six
points cannot constrain, is not tested. Each spectrum keeps
its own normalization. The test
statistic $\dchi^2 = \chi^2(H_0) - \chi^2(H_1)$ is compared with a $\chi^2$
distribution with two degrees of freedom. Pseudo-experiments under $H_0$
confirm this: the observed 0--5\% value is exceeded in $1.7 \pm 0.3\%$
($0.4 \pm 0.2\%$) of 1500 (1000) trials with quadrature (correlated)
uncertainties, against 1.7\% (0.2\%) asymptotically. Significances are quoted as
one-sided Gaussian equivalents of the resulting $p$ value. The joint test is
our primary statistic.

\section{Bulk reference}
\label{sec:bulk}

A simultaneous fit to $\pi^\pm$, $K^\pm$, $p$ and $\bar p$ in the ALICE
default ranges ($\pi$: 0.5--1.0, $K$: 0.2--1.5, $p$: 0.3--3.0~GeV/$c$;
130 points) gives
\begin{gather*}
\Tkin = 94.9 \pm 3.6~\text{MeV},\\
\betaavg = 0.651 \pm 0.004,\quad
n = 0.712 \pm 0.019,
\end{gather*}
with $\chi^2/\text{ndf} = 17.7/121$, small because largely
$\pT$-correlated systematic uncertainties are added to each point, as in
the ALICE fit. The uncertainties are fit
uncertainties obtained from the stat$\,\oplus\,$syst point uncertainties.
They correspond to the first uncertainty quoted by ALICE, which also
includes the bin-by-bin systematic uncertainties. The correlation
coefficients are $\rho(T,\betaavg) = -0.50$, $\rho(T,n) = -0.15$ and
$\rho(\betaavg,n) = -0.74$. Our values reproduce the ALICE
result~\cite{ALICE:2013mez} ($\Tkin = 95 \pm 4 \pm 10$~MeV,
$\betaavg = 0.651 \pm 0.004 \pm 0.020$, $n = 0.712 \pm 0.019 \pm 0.086$),
which validates the implementation (Appendix~\ref{app:validation}). The
ALICE fit-range and species variations, which make up most of the ALICE
second (systematic) uncertainty, are propagated explicitly in
Sec.~\ref{sec:variations}.

\section{The \texorpdfstring{$\phi$}{phi} in 0--5\% central collisions}
\label{sec:phi}

\subsection{Baseline result}

\begin{figure}[t]
\centering
\includegraphics[width=\columnwidth]{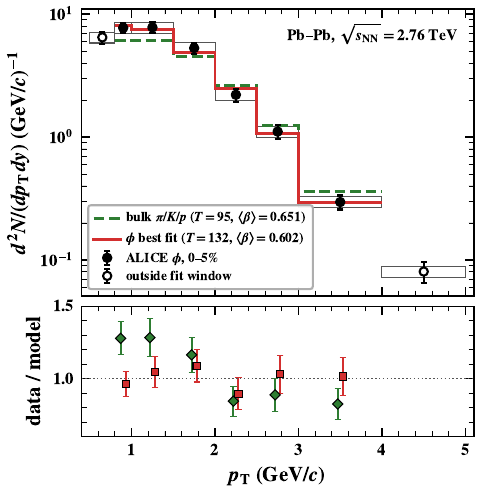}
\caption{ALICE $\phi(1020)$ spectrum in 0--5\% Pb--Pb collisions at
$\sqrts = 2.76$~TeV~\cite{ALICE:2014jbq_phi} with bin-averaged BGBW
curves for the bulk point ($T = 94.9$~MeV, $\betaavg = 0.651$, dashed)
and the $\phi$ best fit ($T = 132$~MeV, $\betaavg = 0.602$, solid), both at
$n = 0.712$ with free normalization. Open points lie outside the fit
window. Lower panel: data/model. Bars show the statistical and
$\pT$-uncorrelated uncertainties; boxes show the 10\% $\pT$-correlated
uncertainty.\label{fig:spectrum}}
\end{figure}

\begin{figure}[t]
\centering
\includegraphics[width=\columnwidth]{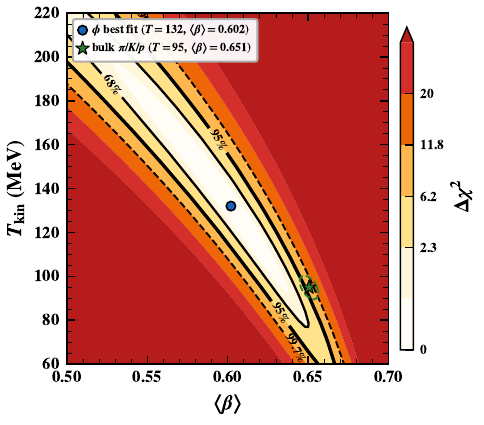}
\caption{$\dchi^2$ of the $\phi$ fit in the $(T, \betaavg)$ plane at
$n = 0.712$, quadrature covariance. Contours mark $\dchi^2 = 2.30$, 6.18
and 11.8 (68.3\%, 95.4\% and 99.7\% confidence level for two parameters). The star is the bulk point
with its $1\sigma$ ($\dchi^2 = 2.30$) ellipse from the bulk fit. The
contours extend beyond the plotted range along the $T$--$\betaavg$
valley.\label{fig:contour}}
\end{figure}

Figure~\ref{fig:spectrum} shows the $\phi$ spectrum with the bulk
prediction and the $\phi$ best fit. Figure~\ref{fig:contour} shows the
$\dchi^2$ landscape in the $(T, \betaavg)$ plane at $n = 0.712$. With the
baseline covariance, the $\phi$ best fit is at
$(T, \betaavg) = (132~\text{MeV}, 0.602)$ with
$\chi^2_{\rm min}/\text{ndf} = 1.09/3$. The one-dimensional profile
intervals ($\dchi^2 = 1$) are $93 < T_\phi < 197$~MeV and
$0.530 < \betaavg_\phi < 0.638$. The temperature is poorly constrained
along the usual $T$--$\betaavg$ anticorrelation valley. The flow velocity,
however, lies below the bulk value $0.651$. The bulk point lies at
$\dchi^2_{\rm fp} = 8.5$ ($2.2\sigma$). In the joint nested test, the bulk
parameters can adjust within their own uncertainties, and
$\dchi^2 = 8.1$ ($p = 0.017$, $2.1\sigma$).

\subsection{Correlation model of the \texorpdfstring{$\phi$}{phi} systematic uncertainty}

Table~\ref{tab:cov} lists the fixed-point $\dchi^2$ for the covariance
models of Sec.~\ref{sec:cov}. A fully correlated 10\% scale uncertainty
gives the largest separation, $\dchi^2 = 13.4$ ($3.0\sigma$). If the same
uncertainty can instead tilt the spectrum, or is correlated only over
0.5--1~GeV/$c$, $\dchi^2$ drops to $5.3$--$6.0$
($1.5$--$1.7\sigma$). Without detailed knowledge of the sources of the
correlated systematic uncertainty, this spread is an irreducible
uncertainty of the test. The quadrature baseline lies in the middle of
the range.

\begin{table}[t]
\caption{Fixed-point $\dchi^2$ (two parameters) at the bulk point for
different treatments of the 10\% $\pT$-correlated $\phi$ uncertainty,
0--5\% centrality.\label{tab:cov}}
\begin{ruledtabular}
\begin{tabular}{l c c}
Covariance model & $\dchi^2$ & Significance \Tstrut\Bstrut\\
\hline
Quadrature (baseline)            & 8.48  & $2.2\sigma$ \Tstrut\\
Smooth, $L = 0.5$~GeV/$c$         & 5.98  & $1.6\sigma$ \\
Smooth, $L = 1.0$~GeV/$c$         & 6.04  & $1.7\sigma$ \\
Smooth, $L = 2.0$~GeV/$c$         & 8.54  & $2.2\sigma$ \\
Linear tilt                      & 5.32  & $1.5\sigma$ \\
Fully correlated (scale)         & 13.36 & $3.0\sigma$ \\
Correlated part omitted (reference only) & 16.12 & $3.4\sigma$ \Bstrut\\
\end{tabular}
\end{ruledtabular}
\end{table}

\subsection{Bulk fit-range and species variations}
\label{sec:variations}

ALICE assessed the systematic uncertainty of its blast-wave parameters by
varying the fit ranges and by excluding individual
species~\cite{ALICE:2013mez}. We repeat these variations and refit the bulk
each time. Where the bulk $\pT$ window changes, we also apply a matching
$\phi$ window. The results are listed in Table~\ref{tab:variations}. The
fit-range variations move the bulk temperature between 83 and 103~MeV but
leave $\betaavg$ within $0.650$--$0.657$. Only the exclusion of the protons
moves the flow velocity appreciably, to $0.631 \pm 0.009$. In that case
the joint test gives $\dchi^2 = 3.3$ ($0.9\sigma$): the $\pi$, $K$ and
$\phi$ spectra are described well by a common freeze-out. The $\phi$--bulk
tension is therefore essentially a $\phi$--proton tension. In the low-$\pT$
variation the bulk temperature is poorly constrained
($89.6 \pm 8.6$~MeV), and the joint test, which can exploit this freedom,
gives $\dchi^2 = 4.9$--$5.4$, well below the fixed-point value.

Across all variations the joint test spans $\dchi^2 = 3.3$--$8.4$
($0.9$--$2.2\sigma$) with the quadrature covariance, and $3.8$--$13.1$
($1.0$--$3.0\sigma$) with the fully correlated one.

\begin{table*}[t]
\caption{Bulk fit-range and species variations following
Ref.~\cite{ALICE:2013mez}. For each variation the bulk is refitted
(both charges, stat$\,\oplus\,$syst), and the fixed-point and joint
$\dchi^2$ are given for the quadrature and fully correlated $\phi$
covariance. ``Matched'' uses a $\phi$ window that follows the bulk window
(high-$\pT$: bins 1.0--5.0; low-$\pT$: bins 0.5--3.0~GeV/$c$). In the
high-$\pT$ matched case the $\phi$ alone does not constrain $(T, \betaavg)$:
its minimum lies at $T > 300$~MeV.\label{tab:variations}}
\begin{ruledtabular}
\begin{tabular}{l c c c c c c c}
& \multicolumn{3}{c}{Bulk fit} & \multicolumn{2}{c}{Quadrature} & \multicolumn{2}{c}{Correlated} \\
Variation & $\Tkin$ (MeV) & $\betaavg$ & $n$ & fixed & joint & fixed & joint \Tstrut\Bstrut\\
\hline
Default                & $94.9 \pm 3.6$  & 0.651 & 0.712 & 8.5  & 8.1 & 13.4 & 12.7 \Tstrut\\
High-$\pT$ ranges      & $102.8 \pm 4.6$ & 0.650 & 0.680 & 7.2  & 6.8 & 11.6 & 10.7 \\
\quad matched $\phi$   &                 &       &       & 6.6  & 6.5 & 8.9  & 8.6 \\
Low-$\pT$ ranges       & $89.6 \pm 8.6$  & 0.654 & 0.734 & 14.0 & 4.9 & 20.3 & 6.1 \\
\quad matched $\phi$   &                 &       &       & 11.8 & 5.4 & 17.0 & 6.7 \\
$\pi$ from 0.3~GeV/$c$ & $91.7 \pm 2.6$  & 0.652 & 0.717 & 8.6  & 8.2 & 13.6 & 12.9 \\
$\pi$ from 0.4~GeV/$c$ & $93.7 \pm 3.1$  & 0.651 & 0.714 & 8.5  & 8.1 & 13.4 & 12.7 \\
Without $\pi$          & $82.4 \pm 7.4$  & 0.657 & 0.721 & 8.7  & 8.4 & 13.8 & 13.1 \\
Without $K$            & $96.9 \pm 3.9$  & 0.656 & 0.680 & 8.3  & 7.9 & 13.1 & 12.4 \\
Without $p$            & $99.7 \pm 6.9$  & 0.631 & 0.802 & 9.7  & 3.3 & 14.7 & 3.8 \Bstrut\\
\end{tabular}
\end{ruledtabular}
\end{table*}

\subsection{Further checks}

\emph{Leave-one-out.} Dropping each of the six $\phi$ points in turn gives
$\dchi^2_{\rm fp} = 5.8$--$7.8$ (quadrature) and $9.5$--$13.0$
(correlated). No single point drives the result.

\emph{$\phi$ fit window.} All contiguous windows with at least five of the
eight published points give $\dchi^2_{\rm fp} = 4.6$--$10.6$
($1.3$--$2.6\sigma$, quadrature). Windows that include the 4--5~GeV/$c$ bin
have a markedly worse $\chi^2_{\rm min}$ and a $\phi$ minimum at high
temperature, consistent with the onset of non-hydrodynamic
production~\cite{ALICE:2017ban}.

\emph{Bin averaging.} Evaluating the model at bin centers instead of
averaging over the bins lowers the baseline to $\dchi^2_{\rm fp} = 5.8$,
because the curvature of the spectrum within the wide bins is then
neglected. All results use bin averaging.

\emph{Flow profile.} At fixed bulk $(T, \betaavg)$, the fixed-point
$\dchi^2$ depends strongly on $n$, and $n$ is strongly anticorrelated with
$\betaavg$ in the bulk fit ($\rho = -0.74$). For this reason $n$ is a free
common parameter in the joint test rather than a scanned nuisance
parameter.

\subsection{Consistency with the \texorpdfstring{$p/\phi$}{p/phi} ratio}
\label{sec:pphi}

\begin{figure}[t]
\centering
\includegraphics[width=\columnwidth]{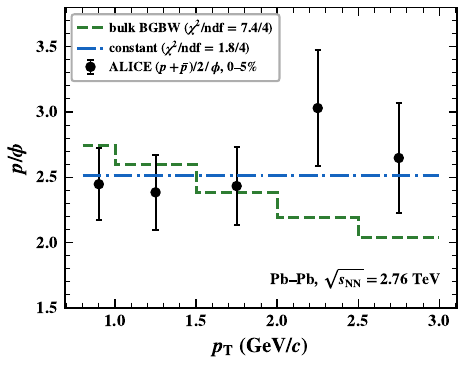}
\caption{$(p + \bar p)/2$ to $\phi$ ratio in 0--5\% Pb--Pb collisions,
computed from the ALICE spectra~\cite{ALICE:2013mez,ALICE:2014jbq_phi} in
the $\phi$ bins, compared with a constant and with the bulk-BGBW
expectation (both normalized to the data). Uncertainties: proton
stat$\,\oplus\,$syst and $\phi$ statistical and $\pT$-uncorrelated.\label{fig:pphi}}
\end{figure}

The bulk blast-wave describes the proton spectrum. Because the $\phi$ and
proton masses differ by 8\%, the bulk parameters predict a $p/\phi$ ratio
that falls by about 25\% between 0.9 and 2.75~GeV/$c$ (Fig.~\ref{fig:pphi}).
The measured ratio, $(p + \bar p)/2$ averaged into the five $\phi$ bins
in 0.8--3.0~GeV/$c$ (within the proton fit range) at 0--5\%, is consistent with a constant [$\chi^2/\text{ndf} = 1.8/4$].
The bulk-BGBW shape gives $\chi^2/\text{ndf} = 7.4/4$ ($p = 0.12$).
A linear fit gives a relative slope of $+0.08 \pm 0.10$ per GeV/$c$,
compared with $-0.16$ per GeV/$c$ for the bulk prediction, a $2.6\sigma$
difference ($2.1\sigma$ if all $\phi$ uncertainties are included). The
flat $p/\phi$ ratio reported by ALICE~\cite{ALICE:2014jbq_phi} is
therefore not in conflict with the present result. It is the same
information viewed differently: a flat ratio between two species of
nearly equal mass requires a $\phi$ spectrum slightly softer than the
bulk BGBW prediction, which is what the fit finds. Neither observable
alone excludes common freeze-out at the current precision.

\subsection{\texorpdfstring{$\phi$}{phi} regeneration}
\label{sec:regen}

Part of the measured $\phi$ yield may come from regeneration in the
hadronic phase ($K\bar K \to \phi$ and other channels). Such $\phi$ would
carry momenta characteristic of the late, cooler stage. We model the
measured spectrum as a mixture,
\begin{equation}
y_i = A\left[(1-f)\,S_i(T_\phi, \betaavg_\phi) + f\,S^{\rm reg}_i\right],
\end{equation}
with shapes $S$ normalized to unit yield in the fit window and regenerated
fraction $f$. The fit returns the primordial parameters. The null
hypothesis places the primordial component at the bulk point; its
spectrum is then the bulk shape only if $S^{\rm reg}$ also has the bulk
shape. For $S^{\rm reg}$ at the bulk point and
$f$ up to $0.29$ (the SMASH box of Sec.~\ref{sec:box} gives 0.19--0.33
among the alive $\phi$ after 10--20~fm/$c$), $\dchi^2_{\rm fp}$ changes only from
8.5 to 8.4, while the primordial component moves to smaller flow
($\betaavg_\phi = 0.589$). A regenerated component softer than the bulk
(bulk flow, $T$ lowered by 20~MeV) softens the null-hypothesis spectrum
itself and brings it closer to the data: $\dchi^2_{\rm fp}$ decreases to
6.6 at $f = 0.15$ and to 5.0 at $f = 0.29$.

The hybrid calculation of Sec.~\ref{sec:hybrid} provides both inputs for
the experiment-matched $\phi$. The regenerated fraction is $f = 0.26$ of
the yield in the fit window, falling from 0.31 to 0.19 across it, and the
regenerated spectrum is softer than the bulk expectation. With this
spectrum as $S^{\rm reg}$ and $f = 0.26$, $\dchi^2_{\rm fp}$ falls to 2.9
(quadrature) and 5.0 (correlated), i.e.\ $0.7\sigma$ and $1.4\sigma$. If
only its distortion relative to the hybrid bulk is transferred to the data
bulk point, the values are 2.0 and 3.3. A regenerated component of the size and
shape predicted by SMASH thus accounts for most of the observed tension,
with the primordial $\phi$ at the bulk point. A regenerated component
with the bulk shape cannot produce the tension, but a soft one can. The
measured $\phi$--bulk difference is therefore compatible both with an early
decoupling of the primordial $\phi$ and with soft regeneration at a common
freeze-out, and the spectra alone do not separate the two.

\section{Centrality dependence}
\label{sec:cent}

\begin{figure}[t]
\centering
\includegraphics[width=\columnwidth]{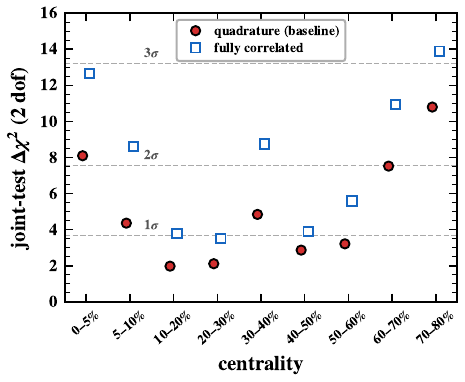}
\caption{Joint-test $\dchi^2$ versus centrality for the quadrature (filled)
and fully correlated (open) $\phi$ covariance. Horizontal lines mark
$1\sigma$, $2\sigma$ and $3\sigma$ for two degrees of freedom
(one-sided Gaussian equivalent).\label{fig:cent}}
\end{figure}

If the $\phi$ decouples earlier because of its small cross section, the
effect should be largest where the hadronic phase lasts longest, that is,
in the most central collisions, and it should weaken toward peripheral
collisions. We repeat the full analysis in all nine centrality classes
common to both measurements. The bulk fit is redone in each class with the
default ranges (Table~\ref{tab:cent}, Fig.~\ref{fig:cent}).

The joint-test $\dchi^2$ is $8.1$ (0--5\%), $4.4$ (5--10\%), and about 2
in 10--20\% and 20--30\%. It rises again to 7.5 and 10.8 in the two most
peripheral classes. In 30--50\% the $\phi$ alone does not constrain the
blast-wave parameters (its minimum lies at $T > 300$~MeV). In peripheral
collisions the bulk fit requires $n \gtrsim 1.3$, and the blast-wave
description of a short-lived, small system is itself questionable. We do
not interpret the peripheral values as evidence of early $\phi$
decoupling. The centrality dependence therefore does not show the
systematic trend expected from a hadronic-phase effect. The 0--5\% result
stands as an isolated $\sim 2\sigma$ tension, not as part of a trend.

We do not combine centrality classes. The $\phi$ systematic uncertainties
are largely common to all classes, and a sum of $\dchi^2$ values that
treats them as independent would overstate the combined significance.

\begin{table}[t]
\caption{Bulk fit and joint-test $\dchi^2$ (two parameters) in each
centrality class. Quadrature and fully correlated $\phi$ covariance.\label{tab:cent}}
\begin{ruledtabular}
\begin{tabular}{l c c c c c}
Centrality & $\Tkin$ (MeV) & $\betaavg$ & $n$ & quad. & corr. \Tstrut\Bstrut\\
\hline
0--5\%   & $94.9 \pm 3.6$  & 0.651 & 0.71 & 8.1  & 12.7 \Tstrut\\
5--10\%  & $96.7 \pm 3.5$  & 0.646 & 0.72 & 4.4  & 8.6 \\
10--20\% & $98.8 \pm 3.5$  & 0.639 & 0.74 & 2.0  & 3.8 \\
20--30\% & $101.2 \pm 3.6$ & 0.625 & 0.78 & 2.1  & 3.5 \\
30--40\% & $105.8 \pm 3.7$ & 0.604 & 0.84 & 4.9  & 8.8 \\
40--50\% & $111.4 \pm 4.0$ & 0.574 & 0.95 & 2.9  & 3.9 \\
50--60\% & $118.4 \pm 4.3$ & 0.535 & 1.10 & 3.2  & 5.6 \\
60--70\% & $128.7 \pm 4.7$ & 0.489 & 1.29 & 7.5  & 10.9 \\
70--80\% & $138.5 \pm 5.2$ & 0.438 & 1.58 & 10.8 & 13.9 \Bstrut\\
\end{tabular}
\end{ruledtabular}
\end{table}

\section{Transport-model studies}
\label{sec:transport}

\subsection{Rescattering in a hadron gas: SMASH box}
\label{sec:box}

The box is a cube of side $L = 20$~fm with periodic boundaries, initialized
with thermal momenta at $T = 156$~MeV and the multiplicities of stable
hadrons and $\phi$ only (800 of each pion charge, 140 $K^\pm$, 50 $p$ and
$\bar p$, hyperons, and 20 $\phi$). All hadronic interactions of
SMASH~3.3~\cite{Weil:2016SMASH} are enabled. We analyze 300 events.

Slow $\phi$ decay first in the box frame, so any population selected by
survival is biased toward high momentum. Conversely, the $\phi$ that decay
at a given time are softer than the thermal population. We quantify this with
a toy Monte Carlo: $3\times10^7$ non-interacting $\phi$ with thermal momenta
at $T = 156$~MeV, decaying with lifetime $\gamma\tau$. The toy is analyzed
with the same estimator as the box, a Boltzmann fit
$A\,\mT K_1(\mT/T)$ to all particles (exact for a static source; applied
with a rapidity cut, the same form is biased by several MeV) over
$0.5 < \pT < 3$~GeV/$c$ with bin-averaged model values. At $t = 0$ it
returns $T = 156.01 \pm 0.03$~MeV. For the $\phi$ still alive, the apparent
temperature rises to 158.4~MeV at $t = 10$~fm/$c$ and to 160.7~MeV at
$t = 20$~fm/$c$. The $\phi$ decaying at $t \lesssim 20$~fm/$c$ give
141--146~MeV. Selection by survival alone thus mimics a
2--5~MeV ``heating'' over the hadronic-phase time scale. This is not
negligible compared with the $\phi$--bulk temperature differences one can
expect in the box, so box results are reported both without and with an
inverse-survival weight.

\begin{table*}
\caption{Effective temperatures (MeV) in the SMASH box at time $t$, from
the all-rapidity Boltzmann fit over $0.5 < \pT < 3$~GeV/$c$. The $\phi$
columns refer to the $\phi$ alive at $t$: all of them, all with the
inverse-survival weight (corr.), the primordial ones without and with the
weight, and the regenerated ones. $f_{\rm regen}$ is the regenerated
fraction among the alive $\phi$. The last column is the toy expectation for
alive, non-interacting $\phi$ in a bath held at $T = 156$~MeV; with the
weight, the toy returns 156~MeV at every $t$. Uncertainties are
statistical (bootstrap over events for the weighted columns).\label{tab:box}}
\begin{ruledtabular}
\begin{tabular}{rcccccccccc}
$t$ (fm/$c$) & $\pi$ & $K$ & $p$ & $\phi$ & $\phi$ corr. & $\phi$ prim. &
$\phi$ prim.\ corr. & $\phi$ regen. & $f_{\rm regen}$ & toy \\
\hline
  5 & 159.0(5) & 160.7(10) & 154.7(20) & 153.7(24) & 152.7(22) & 156.1(26)  & 155.0(24)  & 123.5(66) & 0.106(4) & -- \\
 10 & 154.1(5) & 159.8(11) & 149.2(21) & 156.6(25) & 154.7(24) & 160.8(28)  & 158.3(26)  & 126.8(53) & 0.186(5) & 158.4 \\
 15 & 150.4(5) & 155.0(11) & 148.7(22) & 156.3(26) & 153.8(28) & 163.9(31)  & 160.0(33)  & 129.3(44) & 0.265(6) & -- \\
 20 & 148.6(6) & 156.2(11) & 146.4(21) & 152.9(26) & 150.1(27) & 160.7(33)  & 155.9(34)  & 132.5(40) & 0.330(7) & 160.7 \\
 30 & 144.4(6) & 148.2(10) & 145.9(22) & 149.1(26) & 145.5(28) & 160.6(38)  & 153.3(43)  & 135.0(37) & 0.462(7) & -- \\
 50 & 141.8(6) & 145.1(10) & 141.8(20) & 149.0(28) & 143.9(27) & 168.3(53)  & 155.6(54)  & 138.5(34) & 0.660(7) & 168.1 \\
100 & 138.5(6) & 141.1(10) & 139.5(20) & 136.3(27) & 132.2(36) & 168.4(106) & 144.2(109) & 132.3(28) & 0.903(5) & 180.9 \\
\end{tabular}
\end{ruledtabular}
\end{table*}

\begin{figure}
\centering
\includegraphics[width=\columnwidth]{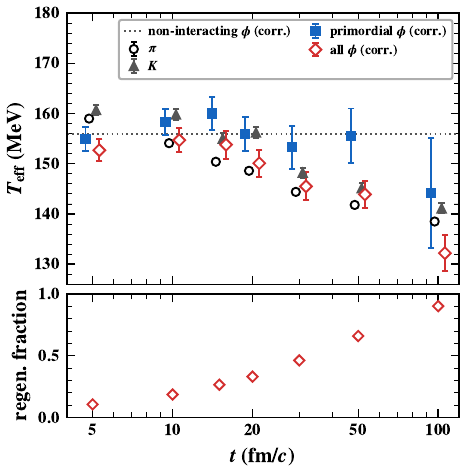}
\caption{Top: effective temperatures in the SMASH box versus time for
pions, kaons, and the alive $\phi$ with the inverse-survival weight, for
primordial $\phi$ only and for all $\phi$. The dotted line is the
non-interacting expectation for the weighted $\phi$. Bottom: regenerated
fraction among the alive $\phi$. Values at different $t$ share particles
and are correlated.\label{fig:box}}
\end{figure}

The results are listed in Table~\ref{tab:box} and shown in
Fig.~\ref{fig:box}. A second, independent production that ends at
20~fm/$c$ agrees within $2\sigma$ at every common time. Four features stand
out.

\emph{The medium cools.} The initial state contains no resonances. As
$\rho$, $K^*$, $\Delta$, \dots\ are populated, kinetic energy goes into
mass, and the pion temperature falls from 159.0~MeV at $t = 5$~fm/$c$ to
141.8~MeV at 50~fm/$c$ and 138.5~MeV at 100~fm/$c$. Kaons and protons
follow. While the system relaxes, the single-species temperatures of pions
and kaons differ by up to $7.6 \pm 1.3$~MeV, although the box has neither
flow nor freeze-out.

\emph{The surviving primordial $\phi$ keep their initial temperature.} With the
inverse-survival weight, the primordial $\phi$ give $155.0 \pm 2.4$~MeV at
5~fm/$c$ and $155.6 \pm 5.4$~MeV at 50~fm/$c$, consistent with the initial
156~MeV, while the pions cool by 17~MeV. At 50~fm/$c$ they lie
$13.8 \pm 5.4$~MeV above the pions and $10.5 \pm 5.5$~MeV above the kaons.
Without the weight, they follow the non-interacting toy
($168.3 \pm 5.3$ against 168.1~MeV at 50~fm/$c$). Two direct measures
confirm that elastic rescattering does little to them. First, among
primordial $\phi$ of the same age, those that have interacted and those
that have not give the same temperature at every time; the largest
difference is $0.9\sigma$, for example $+3.7 \pm 8.5$~MeV at 20~fm/$c$ with
the weight. (Splitting \emph{all} alive $\phi$ by interaction count
instead separates young, regenerated $\phi$ from old, survival-selected
ones and produces a spurious gap of 22~MeV.) Second, the mean change of
$\mT$ in a $\phi$ elastic collision is $+0.6 \pm 2.9$ and
$-0.2 \pm 3.1$~MeV in the windows 0--10 and 10--20~fm/$c$, and only
$-4.3 \pm 1.5$~MeV in 50--100~fm/$c$, when the bath is near 140~MeV. With
1.7 elastic collisions per $\phi$ in 100~fm/$c$, elastic rescattering
cannot keep the $\phi$ in kinetic equilibrium with a cooling medium. In
this controlled setting the box illustrates the premise behind an early
kinetic decoupling of the $\phi$.

The weight corrects for decay but not for absorption. Of the 6000
primordial $\phi$, $34.2 \pm 0.6\%$ are absorbed in inelastic reactions,
$59.4\%$ decay and $6.4\%$ are still alive at 100~fm/$c$. Absorption
favors soft $\phi$: its probability is $35.5 \pm 0.7\%$ for an initial
$\pT < 0.8$~GeV/$c$ and $29.8 \pm 1.3\%$ for 0.8--1.5~GeV/$c$, and the
initial momenta of the $\phi$ that are later absorbed give
$140.8 \pm 4.2$~MeV, against $157.1 \pm 3.3$~MeV for the others (a
difference of $16.3 \pm 5.6$~MeV). The weighted primordial temperature
therefore describes the $\phi$ that escape absorption. They start about
4~MeV above the full primordial sample ($153.1 \pm 2.4$~MeV) and lose
little of this temperature over 50~fm/$c$. Adding back the absorbed $\phi$,
with their momenta just before absorption and decay-weighted up to that
time, gives $144.1 \pm 2.9$ and $141.3 \pm 3.2$~MeV at 30 and 50~fm/$c$,
close to the pions. This combination is not the temperature of any
population present at time $t$, however: just before absorption the
absorbed $\phi$ are soft ($\approx 112$~MeV). Absorption thus acts as a
momentum-dependent filter that hardens the surviving primordial $\phi$, a
selection in the same direction as the time dilation.

\emph{Regeneration ties the $\phi$ population to the medium.} The
regenerated fraction among the alive $\phi$ grows from 11\% at 5~fm/$c$ to
33\% at 20~fm/$c$, 66\% at 50~fm/$c$ and 90\% at 100~fm/$c$. Regenerated
$\phi$ are soft when they are formed. For $K\bar K \to \phi$, which accounts
for 30--45\% of them, this is the detailed-balance factor: in equilibrium
the formation rate at given momentum equals the decay rate, so freshly
formed $\phi$ carry the same $m/E$ weight as the decaying ones. (Over
100~fm/$c$, $K\bar K$ fusion produces 40\% of the regenerated $\phi$,
$\pi\rho \to \phi$ 14\%, and decays of heavier resonances, mostly
$f_2(2010)$, $f_2(2340)$ and $K_4^*(2045)$, 46\%.) Weighted by
$E/m$, the fused $\phi$ reproduce the kaon temperature within $1\sigma$ in
every time window (for example $137.7 \pm 4.0$ against
$141.1 \pm 1.0$~MeV for 50--100~fm/$c$). The $\phi$ from decays of heavier
resonances (40--55\%) are softer still. Because of regeneration, the
weighted temperature of all alive $\phi$ follows the medium
($143.9 \pm 2.7$~MeV at 50~fm/$c$, $2.1 \pm 2.7$~MeV above the pions). The
$\phi$ population as a whole is thus held near the medium temperature by
its chemistry, not by elastic collisions.

\emph{Every selection of a $\phi$ population is biased.} The decaying
$\phi$ (132.8 $\pm$ 2.6~MeV for $\phi \to K^+K^-$ in 0--100~fm/$c$), the
regenerated ones and the survivors are each shifted by the time-dilation
factor, in either direction and by several MeV. The reconstructable decays
are not a physical observable in a box. Daughter kaons interact again
within a few fm/$c$, mostly through $K\pi \to K^*$. The fraction of
$\phi \to K^+K^-$ decays with both kaons free at the end of the run is
$70 \pm 5\%$ for decays less than 2~fm/$c$ before the end, $22 \pm 4\%$ for
2--5~fm/$c$, $2.4 \pm 1.0\%$ for 5--10~fm/$c$, and zero beyond, the same in
both productions. It measures the time left in the run, not a decoupling.
The $\phi$ forced to decay at the end simply reproduce the alive population
($136.1 \pm 2.7$ against $136.3 \pm 2.7$~MeV). We therefore do not use pair
reconstruction in the box.

In summary, the box separates two mechanisms. Elastic rescattering is too
weak to cool the primordial $\phi$ along with the medium. Absorption and
regeneration replace them, over time, with $\phi$ that reflect the current
medium; absorption also removes the softer primordial $\phi$ first.
Whether the measured $\phi$ retain an earlier, hotter or less-flowing
freeze-out therefore depends on how these two mechanisms compete in an
expanding system, as seen through the experimental $\phi$ definition. The
static box, where regeneration has unlimited time at fixed density, cannot
answer this. The hybrid calculation addresses it.

\subsection{vHLLE+SMASH hybrid}
\label{sec:hybrid}

Hybrid approaches describe the dense stage with viscous hydrodynamics
and the dilute hadronic stage with a transport
code~\cite{Petersen:2008dd,Song:2013qma}. We use the vHLLE+SMASH
combination~\cite{Schafer:2021csj} and simulate 0--5\% central Pb--Pb
collisions at $\sqrts = 2.76$~TeV.
The initial entropy density is built from 1000 GLISSANDO
events~\cite{Broniowski:2007nz,Rybczynski:2013yba} of the 0--5\% class,
whose sources are averaged into one smooth profile. The sources are a
mixture of wounded nucleons and binary collisions ($\alpha = 0.15$),
smeared with Gaussians of width 0.4~fm, with a tilted longitudinal
profile~\cite{Bozek:2010bi}. The normalization is set to reproduce the
measured charged-particle density, which the calculation gives as
$dN_{\rm ch}/d\eta = 1569$, against 1601 measured. This averaged initial
state is evolved with the 3+1D viscous hydrodynamics code
vHLLE~\cite{Karpenko:2013wva} from $\tau_0 = 0.6$~fm/$c$. We use a
constant $\eta/s = 0.08$, no bulk viscosity, and the chiral-model
equation of state~\cite{Steinheimer:2010ib}. The particlization surface
at $e_{\rm crit} = 0.306$~GeV/fm$^3$ is constructed with
Cornelius~\cite{Huovinen:2012is}, using the SMASH hadron-gas equation of
state on the surface. Its mean temperature is $154.4 \pm 4.6$~MeV. Hadrons
are sampled on this surface with Grad 14-moment shear corrections and
propagated with SMASH~3.3~\cite{Weil:2016SMASH} to 100~fm/$c$, with weak
decays switched off as in the measurement. Because a single averaged
hydrodynamic event is sampled repeatedly, the calculation contains no
event-by-event fluctuations of the initial geometry. We
generate 50\,000 events with all hadronic interactions and, as a
reference, 10\,000 events in which the hadrons only decay. Every group of
100 events uses an independent random-number stream, and the Monte Carlo
uncertainties are the job-to-job standard deviations.

The $\phi$ is defined as in the measurement: $\phi$ alive at the end of the
evolution, plus $\phi \to K^+K^-$ decays in which neither kaon interacts
again, divided by the branching ratio. We refer to this as the reconstructed
$\phi$. ALICE corrects its yields with HIJING and GEANT3 simulations,
which have no hadronic phase~\cite{ALICE:2014jbq_phi}, so we compare at the
level of produced particles. Requiring that neither kaon interact is the
strictest criterion; a daughter scattered with small momentum transfer may
leave the pair in the mass peak~\cite{ALICE:2014jbq_phi}. For comparison we use the $\phi$ at particlization and all $\phi$
that decay or survive, with their momenta at the last interaction. We also
split the reconstructed $\phi$ into primordial ones, produced at
particlization or in resonance decays without intermediate interaction, and
regenerated ones. Counting decays in any channel instead of $K^+K^-$ changes
none of the conclusions. The $\phi$ at their last interaction add the
decays lost to rescattering (23\% at 0.8--1.0 and 4\% at 3--4~GeV/$c$).
Keeping any part of these gives $\dchi^2_{\rm fp} = 17.6$--33.9, at least
twice the observed 8.5. In the decays-only run the reconstructed primordial
$\phi$ reproduce the particlization spectrum within 0.3\% below 3~GeV/$c$,
and the $K^+K^-$ fraction of the decays matches the branching ratio in every
$\pT$ bin.

The model spectra are analyzed exactly like the data. Each model bin is
assigned the relative ALICE uncertainties of the corresponding data bin,
with the same decomposition and correlation models, and the Monte Carlo
uncertainty is added in quadrature. Since the model spectra do not
fluctuate, the resulting $\dchi^2$ is the value expected if the data
followed the model, given the ALICE uncertainties.

\begin{table*}
\caption{Blast-wave analysis of the vHLLE+SMASH spectra with ALICE
uncertainties, compared with the data. For each $\phi$ definition: best fit
$(T_\phi, \betaavg_\phi)$ at the bulk $n$, the fixed-point $\dchi^2$ with
respect to the bulk point of the same calculation, and the joint-test
$\dchi^2$ with quadrature and correlated $\phi$ uncertainties. $dN/dy$ is per
event in $|y| < 0.5$. The primordial and regenerated components are not
tested jointly.\label{tab:hybrid}}
\begin{ruledtabular}
\begin{tabular}{llcccccc}
 & & $T$ (MeV) & $\betaavg$ & $dN/dy$ & $\dchi^2_{\rm fp}$ & $\dchi^2_{\rm joint}$ (quad.) & $\dchi^2_{\rm joint}$ (corr.) \\
\hline
Decays only & bulk & $137.6 \pm 5.2$ & 0.593 & & & & \Tstrut\\
 & $\phi$ at particlization & 122.7 & 0.622 & 16.7 & 2.1 & 2.0 & 3.5 \\
 & $\phi$ reconstructed & 122.6 & 0.622 & 17.0 & 1.9 & 1.8 & 3.1 \Bstrut\\
\hline
Full & bulk & $91.3 \pm 3.3$ & 0.685 & & & & \Tstrut\\
 & $\phi$ at particlization & 123.3 & 0.623 & 16.8 & 29.2 & 27.9 & 35.1 \\
 & $\phi$ at last interaction & 115.8 & 0.631 & 20.9 & 30.8 & 29.4 & 36.2 \\
 & $\phi$ reconstructed & 102.9 & 0.651 & 17.7 & 20.9 & 19.8 & 27.1 \\
 & \quad primordial & 95.5 & 0.662 & 12.7 & 14.3 & -- & -- \\
 & \quad regenerated & 115.9 & 0.620 & 4.9 & 46.5 & -- & -- \Bstrut\\
\hline
Data & bulk & $94.9 \pm 3.6$ & 0.651 & & & & \Tstrut\\
 & $\phi$ & 132 & 0.602 & & 8.5 & 8.1 & 12.7 \\
\end{tabular}
\end{ruledtabular}
\end{table*}

The results are listed in Table~\ref{tab:hybrid}. Without hadronic
interactions, every $\phi$ definition is consistent with the bulk of the
same calculation (joint $\dchi^2 \le 2.0$ with quadrature and $\le 3.5$
with correlated uncertainties, all below $1\sigma$). The analysis therefore
does not produce a $\phi$--bulk separation when all species leave the
system together. The hadronic phase cools the bulk from 137.6 to 91.3~MeV
and raises its flow from 0.593 to 0.685, comparable to the values found in data.
The reconstructed $\phi$ follow only part of this evolution. Their best fit,
$(102.9~\text{MeV}, 0.651)$, has a flow velocity $0.034$ below the bulk,
and the joint test gives $\dchi^2 = 19.8$ (quadrature) and 27.1
(correlated). Measured with the ALICE uncertainties, the model's
$\phi$--bulk separation would thus appear at $3.9\sigma$ and $4.7\sigma$.
The Monte Carlo uncertainty of these predictions, from a bootstrap over the
500 independent jobs, is $\pm 0.4$ in $\dchi^2_{\rm fp}$ and $\pm 0.001$ in
the flow difference, which is negligible.

Three effects contribute in the model. First, most primordial $\phi$ do
not rescatter. 72\% of the reconstructed primordial $\phi$ have no
interaction after particlization, which takes place at a mean lab time of
10.8~fm/$c$. Averaged over all of them, the last interaction occurs only
1.1--3.6~fm/$c$ after particlization (the range reflects decayed $\phi$
whose particlization time is not recorded). The final momenta of pions, kaons and protons, by
contrast, are mostly set late. Most of them are produced in resonance
decays during the hadronic phase, at mean times of 25, 22 and 29~fm/$c$,
and afterwards mostly stream freely. Only 5\%, 17\% and 2\% of them are
primordial hadrons without any interaction in their history. Second, 28\% of the reconstructed
$\phi$ are regenerated, mostly through $K\bar K$ fusion (20\%), and their
spectrum is softer than the bulk expectation and even softer than at
particlization (mean $\pT$ of 1.38~GeV/$c$ against 1.42~GeV/$c$ at
particlization, 1.49~GeV/$c$ for the reconstructed primordial $\phi$, and
1.61~GeV/$c$ expected at the bulk point, all within 0.5--5~GeV/$c$). Regeneration therefore increases the separation from
the bulk. This is the detailed-balance softening found in the box
(Sec.~\ref{sec:box}). Third, the reconstruction selects against soft $\phi$.
A $\phi$ that decays inside the medium loses its daughters to rescattering:
the lost fraction of $K^+K^-$ decays falls from 28\% at
$0.5 < \pT < 0.8$~GeV/$c$ to 6\% at 3--4~GeV/$c$. This hardens the
reconstructed spectrum relative to the $\phi$ at their last interaction and
reduces $\dchi^2_{\rm fp}$ from 30.8 to 20.9. In the model, the measurable
$\phi$--bulk difference is thus a combination of early decoupling,
regeneration, and reconstruction losses, not a clean measure of the
first.

\emph{Comparison with data.} The measured $\phi$ lies between the common
freeze-out and the model prediction. To compare the two, we transfer the
model distortion to the data bulk point bin by bin: the model $\phi$
spectrum divided by the blast-wave at the model bulk point multiplies the
blast-wave at the data bulk point. This removes the differences in bulk
point and flow profile, assuming that the distortion does not depend on
them. The calculation uses one
hydrodynamic setup (a smooth initial state, one $\eta/s$ and one
particlization energy density), whose bulk flow exceeds the measured one by
0.034; variations of this setup are not explored, and the probabilities
below include only the experimental uncertainties. The data are compatible
with the transferred spectrum [$\dchi^2 = 2.4$ ($p = 0.30$) with
quadrature and 3.7 ($p = 0.15$) with correlated uncertainties], while the
common freeze-out point is disfavored at 8.5 and 13.4. The size of the
separation is smaller than predicted. Without fluctuations, the transferred
spectrum gives $\dchi^2_{\rm fp} = \lambda = 20.9$ (quadrature) and 28.2
(correlated) for the data uncertainties. Under a noncentral $\chi^2_2$
distribution with noncentrality $\lambda$, values at or below the observed
8.5 and 13.4 occur with probability 0.04; the model's own separation
(Table~\ref{tab:hybrid}) gives the same. The
data are thus consistent with the direction of the predicted displacement,
and with about two-thirds of its size [$(8.5/20.9)^{1/2} = 0.64$].

\section{Multistrange baryons}
\label{sec:multistrange}

The hybrid calculation shows that, for the $\phi$, early decoupling and
regeneration cannot be separated in the spectrum. The $\Omega$ avoids this
ambiguity: in SMASH no reaction or decay produces an $\Omega$, so its yield
can only decrease in the hadronic phase (to 0.87 of the particlization
yield). We therefore apply the same test to the $\Xi$ and $\Omega$ spectra
measured by ALICE~\cite{ALICE:2014jbq}. The design was fixed before any fit
and is documented with the analysis code. The $\Omega$ is the primary probe.
The $\Xi$ is secondary, because it is fed by $\Xi^*$ decays throughout the
hadronic phase.

The multistrange spectra are available for 0--10\% central collisions. We
form the 0--10\% bulk and $\phi$ spectra as the mean of the 0--5\% and
5--10\% spectra, with statistical uncertainties combined in quadrature and
systematic ones linearly. Particles and antiparticles are averaged. The
$\Xi$ and $\Omega$ spectra are $d^2N/(d\pT\,dy)$, so the model carries the
$2\pi\pT$ Jacobian and is averaged over each bin, as for the $\phi$. The fit
windows cover the same $\mT - m$ range as the $\phi$ window,
0.34--2.63~GeV: bin centers $1.0 < \pT < 3.7$~GeV/$c$ (14 points) for the
$\Xi$ and $1.1 < \pT < 4.0$~GeV/$c$ (9 points) for the $\Omega$. The published
systematic uncertainty is a single total per point, a constant fraction of
the yield (5\% for $\Xi$, 7\% for $\Omega$). We treat it in quadrature
(primary), as fully correlated, and as a linear tilt. A fully correlated
constant fraction is a pure scale uncertainty that the free normalization
absorbs, so the correlated case tests the shape against the statistical
uncertainties alone. The model rows use the 0--5\% calculation of
Sec.~\ref{sec:hybrid}, analyzed with the 0--10\% data uncertainties.

\begin{table*}
\caption{Multistrange baryons in 0--10\% central collisions: best fit
$(T, \betaavg)$ at the bulk $n$ and joint-test $\dchi^2$ against the bulk
point, for the data and for the vHLLE+SMASH calculation with the same
relative uncertainties. ``$\Xi+\Omega$'' is a common fit of both species.
The 0--10\% data bulk gives $T = 95.7 \pm 3.5$~MeV, $\betaavg = 0.648$,
$n = 0.717$; the model bulk gives $(137.6~\text{MeV}, 0.593)$ without and
$(91.4~\text{MeV}, 0.685)$ with hadronic interactions.\label{tab:ms}}
\begin{ruledtabular}
\begin{tabular}{llcccc}
 & & $(T, \betaavg)$ & $\dchi^2_{\rm joint}$ (quad.) & $\dchi^2_{\rm joint}$ (corr.) & $\dchi^2_{\rm joint}$ (tilt) \\
\hline
Data & $\Omega$ & (202, 0.528) & 82.1 & 89.2 & 53.0 \Tstrut\\
 & $\Xi$ & (109, 0.612) & 148.1 & 209.6 & 68.8 \\
 & $\Xi+\Omega$ & (128, 0.597) & 168.7 & 211.9 & 89.7 \\
 & $\phi$ & (152, 0.585) & 6.5 & 12.8 & 5.1 \Bstrut\\
\hline
Model, decays only & $\Omega$ & (121, 0.612) & 1.1 & 1.6 & \Tstrut\\
 & $\Xi$ & (124, 0.609) & 1.4 & 3.5 & \\
 & $\Xi+\Omega$ & (122, 0.610) & 2.2 & 4.2 & \Bstrut\\
\hline
Model, full & $\Omega$ & (90, 0.646) & 103.5 & 109.4 & \Tstrut\\
 & $\Xi$ & (87, 0.662) & 173.1 & 240.2 & \\
 & $\Xi+\Omega$ & (100, 0.651) & 194.1 & 233.7 & \\
 & $\phi$ & (103, 0.651) & 23.4 & 34.6 & \\
\end{tabular}
\end{ruledtabular}
\end{table*}

\begin{figure}
\centering
\includegraphics[width=\columnwidth]{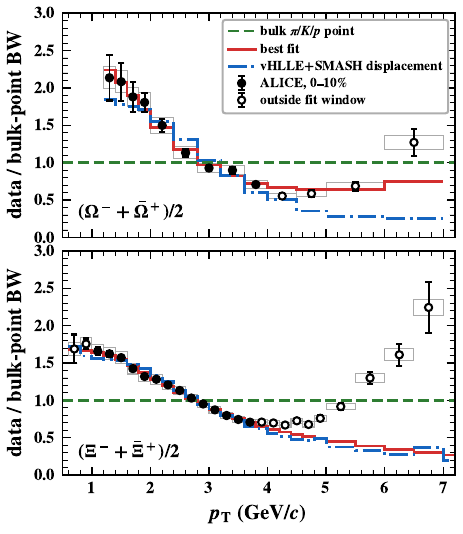}
\caption{Measured $\Omega$ (top) and $\Xi$ (bottom) spectra in 0--10\%
central collisions divided by the blast-wave at the bulk point (dashed line
at unity), each normalized in the fit window. Also shown are the best fit
(solid) and the vHLLE+SMASH distortion transferred bin by bin to the data
bulk point (dash-dotted). Bars are statistical and boxes systematic uncertainties. Open
points lie outside the fit window.\label{fig:ms}}
\end{figure}

The results are listed in Table~\ref{tab:ms} and shown in
Fig.~\ref{fig:ms}. Both multistrange species are far from the bulk
freeze-out, while their own best fits describe them well
($\chi^2_{\rm min}/\text{ndf} = 2.1/6$ for the $\Omega$ and $1.7/11$ for the
$\Xi$ with quadrature uncertainties, 5.7/6 and 14.1/11 with correlated
ones). For the $\Omega$ the joint test gives
$\dchi^2 = 82$ with quadrature, 89 with correlated and 53 with tilted
uncertainties, for two degrees of freedom. The three window variations
fixed in advance (lower edge one bin higher, upper edge one bin lower,
upper edge at 3~GeV/$c$) give 51--73. The $\Xi$ gives 69--210, and
78--137 under the window variations. Every window and uncertainty variant
thus corresponds to more than $6.7\sigma$. (Gaussian equivalents this large
rely on the asymptotic $\chi^2$ distribution far into its tail; we use them
only to show that the separation is not a fluctuation.)

As for the $\phi$, the separation is anchored by the protons. Excluding
pions or kaons from the bulk reference changes the $\Omega$ result to 80 and
76 and the $\Xi$ result to 130 and 132. Without the protons the values fall
to 16.2 ($3.4\sigma$) for the $\Omega$ and 8.8 ($2.2\sigma$) for the $\Xi$.
Pions and kaons alone constrain the flow only weakly, and the common
freeze-out point can then move to accommodate the heavy baryons. This is a
loss of sensitivity rather than of the effect. In the full hybrid
calculation, where the multistrange baryons do lag the bulk flow, the same
variation reduces the separation from 103 and 173 to 6.6 and 6.7
($1.8\sigma$). The measured $\Omega$ retains more than this.
A common fit of $\Xi$ and $\Omega$ gives $(T, \betaavg) = (128~\text{MeV}, 0.597)$. This
agrees with the 0--10\% multistrange parameters of Melo and
Tom\'a\v{s}ik, $(126~\text{MeV}, 0.605)$~\cite{Melo:2015bwa}, obtained with a
different blast-wave implementation that includes resonance decays. The
$\phi$ in the same 0--10\% class gives 5.1--12.8 (1.4--2.9$\sigma$),
consistent with the 0--5\% result.

\emph{The test is valid for heavy baryons.} With 2\% statistical
uncertainties, the $\Xi$ spectrum tests the blast-wave shape far more
precisely than the $\phi$. A common blast-wave might therefore fail for the
$\Xi$ and $\Omega$ even at a common freeze-out, for example because of the
different feed-down. The decays-only calculation rules this out. There,
all species leave the particlization surface together and pass only
through resonance decays, and $\Xi$, $\Omega$ and their common fit agree
with the bulk point ($\dchi^2 \le 4.2$, below $1.2\sigma$). This check was
added to the design after the first results; we report it as such.

\emph{Comparison with transport.} With hadronic interactions, the model
separates $\Xi$ and $\Omega$ from its own bulk by a similar amount
($\dchi^2 = 103$--240). The hadronic phase raises the bulk flow by 0.092,
but that of the $\Omega$ by only 0.034 and that of the $\Xi$ by 0.053.
This is not because the $\Omega$ avoids collisions: 64\% of the final
$\Omega$ interact after particlization, with 1.3 elastic collisions on
average. The bulk flow, by contrast, is carried by pions, kaons and
protons emitted late in resonance decays (Sec.~\ref{sec:hybrid}). Since
nothing produces $\Omega$ in the hadronic phase, the $\Omega$ separation in
the model involves no regeneration. Hybrid calculations with other
hadronic cascades also find that multistrange hadrons scatter less than
non-strange ones in the hadronic stage, so that their spectra reflect an
earlier stage of the expansion~\cite{Takeuchi:2015ana}.

The size of the measured separation agrees with the model. Transferred to
the data bulk point bin by bin, as for the $\phi$ (Sec.~\ref{sec:hybrid}),
the model distortions give $\lambda = 128$ ($\Omega$) and 400 ($\Xi$),
against 97 and 344 observed ($P = 0.06$ and 0.07 for values this low or
lower). In mean transverse momentum within the fit window, the model
accounts for 83\% of the deficit of the measured $\Omega$ with respect to
the bulk expectation (2.13 against 2.09~GeV/$c$ measured and 2.31~GeV/$c$
at the bulk point) and for 91\% of that of the $\Xi$. The detailed shapes
differ. The data against the transferred spectra give $\dchi^2 = 9.1$
($p = 0.011$) for the $\Omega$ and 7.0 ($p = 0.03$) for the $\Xi$ with
quadrature uncertainties, and 20 and 37 with correlated ones. The measured
spectra are more convex, with excesses of 9--17\% at both ends of the
$\Omega$ window relative to the model shape. Non-thermal production by
coalescence is expected to contribute at high $\pT$. In the 0--10\% class
the $\phi$ shows less separation than the model predicts ($\lambda = 25.0$
against 6.8 observed, $P = 0.006$), so the transport calculation describes
the multistrange baryons better than the $\phi$.

\begin{table}
\caption{Multistrange baryons versus centrality, with the same fit windows
and code as in 0--10\%. Bulk point, joint-test $\dchi^2$ for quadrature
(tilted) uncertainties, and the mean $\pT$ in the fit window relative to
the blast-wave at the bulk point. The statistical uncertainty of the latter
is 0.2--1.3\% ($\Omega$) and 0.2--0.4\% ($\Xi$).\label{tab:mscent}}
\begin{ruledtabular}
\begin{tabular}{lccccc}
Class & Bulk $(T, \betaavg)$ & \multicolumn{2}{c}{$\dchi^2_{\rm joint}$} & \multicolumn{2}{c}{$\langle\pT\rangle$ deficit} \\
 & & $\Omega$ & $\Xi$ & $\Omega$ & $\Xi$ \\
\hline
0--10\%  & (95.7, 0.648)  & 82 (53) & 148 (69) & $-9.6\%$ & $-6.7\%$ \Tstrut\\
10--20\% & (98.8, 0.639)  & 70 (40) & 126 (45) & $-8.2\%$ & $-6.8\%$ \\
20--40\% & (103.0, 0.617) & 87 (45) & 101 (34) & $-9.6\%$ & $-6.6\%$ \\
40--60\% & (114.0, 0.559) & 53 (32) & 47 (38)  & $-7.9\%$ & $-4.9\%$ \\
60--80\% & (132.6, 0.472) & 7.1 (9.0) & 15 (22) & $-3.5\%$ & $-3.1\%$ \\
\end{tabular}
\end{ruledtabular}
\end{table}

\emph{Centrality dependence.} We repeat the test in the other four
multistrange classes with the same windows and code. Each bulk reference
is the mean of the two ALICE $\pi/K/p$ classes the class spans
(Table~\ref{tab:mscent}). The $\dchi^2$ values mix the size of the effect
with the precision of the data, so we also give the deficit of the mean
$\pT$ in the fit window relative to the bulk-point expectation, which
measures the size alone. For both species the deficit is constant from
0--10\% to 20--40\% ($\Omega$ 8--10\%, $\Xi$ 6.6--6.8\%). It then decreases
toward peripheral collisions, to 3.5\% and 3.1\% in 60--80\%, where the
separation is $1.9\sigma$ ($\Omega$) and $3.3\sigma$ ($\Xi$). The transferred
model distortion (0--5\% calculation) corresponds to deficits of 7.9\%
($\Omega$) and 6.1\% ($\Xi$). The weakening toward peripheral
collisions, where the hadronic phase is shorter, is what decoupling from
a flowing medium implies. Unlike the $\phi$ tension, which is already
absent in 10--30\% (Sec.~\ref{sec:cent}), the multistrange separation
persists up to 40--60\%.
\emph{Hybrid calculation in 20--40\%.} To test the centrality trend in the
model, we repeated the vHLLE+SMASH calculation of Sec.~\ref{sec:hybrid}
for 20--40\% central collisions (50\,000 events with and 10\,000 without
hadronic rescattering). Only the initial state changes: it is the
event-averaged GLISSANDO profile of that class, with the 0--5\%
normalization kept. The calculation gives $dN_{\rm ch}/d\eta = 486$,
10\% below the measured value of 538, and its particlization surface has
a mean temperature of $153.7$~MeV. The model spectra are analyzed with the
ALICE 20--40\% uncertainties, and the bulk and $\phi$ references are formed
as for the data, from the 20--30\% and 30--40\% classes. In the
decays-only limit, $\Omega$, $\Xi$, $\phi$ and the common multistrange fit
all agree with the model bulk point ($\dchi^2 \le 3.7$). With
rescattering, the model displaces all three species from its bulk by almost
the same flow velocity as in 0--5\%: $\Delta\betaavg = -0.035$ ($\Omega$),
$-0.031$ ($\Xi$) and $-0.033$ ($\phi$). Because the model bulk has a
flatter flow profile in this class ($n = 0.63$, against 0.80 in the data),
an equal $\Delta\betaavg$ does not mean an equal change of the spectrum.
The model's own mean-$\pT$ deficit of the $\phi$ falls from 7.1\% in
0--5\% to 4.5\% in 20--40\%, together with the regenerated fraction (28\%
to 22\%) and the fraction of $K^+K^-$ decays lost to rescattering (23\% to
15\%). A threefold change in multiplicity thus changes the $\phi$ deficit
by 2.6 percentage points, and the 10\% multiplicity deficit of the
calculation is expected to change it by only a few tenths of a point.

For the multistrange baryons the model again matches the data in size.
The transferred distortions give mean-$\pT$ deficits of 8.2\% ($\Omega$)
and 6.1\% ($\Xi$), against 9.6\% and 6.6\% measured. In $\dchi^2_{\rm fp}$,
$\lambda = 150$ ($\Omega$) and 283 ($\Xi$) compare with 108 and 206
observed ($P = 0.03$ and 0.006; 0.12 and 0.20 with correlated
uncertainties). As in 0--10\%, the measured spectra are more convex than
the model shape, which lowers the observed $\dchi^2_{\rm fp}$ relative to
$\lambda$ even where the mean-$\pT$ deficit is as large as predicted. The
model therefore reproduces the size of the multistrange separation and its
near-constancy from central to mid-central collisions.

For the $\phi$ it overpredicts the separation in both classes. In the data
the $\phi$ is consistent with the bulk in 20--40\%
($\dchi^2_{\rm joint} = 2.4$, or 4.5 with correlated uncertainties). The
transferred model distortion gives $\lambda = 10.9$ (17.6), against 2.5
(4.6) observed, so $P = 0.03$ (0.01) for this calculation; in 0--10\% the
values are $\lambda = 25.0$ (36.1) against 6.8 (13.6), $P = 0.006$
(0.008). The predicted mean-$\pT$ deficit is 5.2\%, compared with 2.4\%
measured; in 0--10\% the two values are 6.9\% and 4.8\%. The $\phi$
separation weakens toward mid-central collisions in both the model and the
data, and the data show about half of the predicted separation in both
classes [$(2.5/10.9)^{1/2} = 0.48$ and $(6.8/25.0)^{1/2} = 0.52$]. The
discrepancy between SMASH and the data is therefore specific to the
$\phi$: compared with this calculation, the measured $\phi$ follows the
bulk more closely, while the multistrange baryons follow the model.

The multistrange baryons therefore show, within the blast-wave framework validated by the
decays-only calculation, that species-dependent kinetic decoupling is
present in the data at high significance relative to the protons, with a
magnitude, and a centrality dependence up to 20--40\%, that the hybrid
calculation reproduces. For the $\Omega$, the model attributes it to a weaker coupling
to the late flow, without regeneration. Against this reference, the $\phi$
shows less separation from the bulk than either the multistrange baryons or
the model expectation, in both centrality classes.

\section{Discussion}
\label{sec:discussion}

The analysis gives a consistent picture. In 0--5\% central Pb--Pb
collisions the $\phi$ spectrum prefers a radial flow velocity about 0.05
smaller than the bulk value. The temperature is only weakly constrained.
The preference is at the $2\sigma$ level, and it depends on two inputs
that the public data do not fix: the bin-to-bin correlation of the 10\%
$\pT$-correlated $\phi$ uncertainty, and the inclusion of the protons in
the bulk reference. The tension is visible equally as a non-flat
expectation for $p/\phi$ that the data do not follow. It does not persist
in 10--30\% central collisions. These findings are compatible with the
qualitative indications of
Refs.~\cite{Melo:2015bwa,Song:2019ncm,Lao:2016epa}. They show, however,
that with present uncertainties the blast-wave analysis of published
spectra cannot by itself establish differential freeze-out of the $\phi$.

A smaller flow velocity of the $\phi$ is the natural signature of earlier
decoupling, since flow builds up during the hadronic phase. Its size can be
compared directly with transport. Analyzed like the data,
vHLLE+SMASH predicts a $\phi$ flow velocity 0.034 below that of the bulk.
The data agree in direction, at about two-thirds of the predicted size: the
predicted spectrum is allowed at $p = 0.15$--0.30, and a separation as
small as the observed one occurs in the model with probability 0.04.
In SMASH the small $\phi$ cross section does produce an early decoupling:
most primordial $\phi$ never rescatter after particlization. The model
also shows, however, that this is not the only source of a smaller
apparent flow. Regenerated $\phi$, 28\% of the reconstructed yield, are
softer than the bulk, and their admixture alone can account for most of the
observed tension (Sec.~\ref{sec:regen}). The $\phi$ spectrum therefore tests
the combined effect of decoupling and regeneration, which transport models
must describe together. In 0--10\% and 20--40\% central collisions the
data show about half of the predicted separation, and the prediction of
this calculation is disfavored at $P = 0.006$ and 0.03
(Sec.~\ref{sec:multistrange}).

The multistrange baryons provide the reference that the $\phi$ alone
cannot (Sec.~\ref{sec:multistrange}). In SMASH the $\Omega$ is not
regenerated, yet it gains only about a third of the hadronic-phase flow of
the bulk. The resulting separation from the $\pi/K/p$ freeze-out, anchored
by the protons, is significant in the data, and its size agrees with the
model. Within the blast-wave description, species-dependent kinetic
decoupling is therefore present in central Pb--Pb collisions, as anticipated by Melo
and Tom\'a\v{s}ik~\cite{Melo:2015bwa}, and a hybrid model with standard
hadronic cross sections accounts for its magnitude. Measured against this
benchmark, the $\phi$ is separated from the bulk less than the model
predicts, in 0--10\% ($P = 0.006$) and in 20--40\% ($P = 0.03$), where
the model reproduces the multistrange separation again. In the model, the $\phi$ separation is the sum of three effects
of comparable size: early decoupling, soft regeneration and reconstruction
losses. Any of them could be misdescribed. Observables sensitive to the
regenerated component, such as the centrality dependence of the $\phi$
yield relative to kaons, could identify which one.

Three improvements would sharpen the test: (i) publication of the
bin-to-bin correlation (or the individual sources) of the $\phi$
systematic uncertainties, which currently sets the dominant spread in
significance; (ii) $\phi$ and identified-hadron spectra with common
systematic treatment from the higher-statistics LHC Run~2 and Run~3
samples; and (iii) simultaneous fits of spectra and elliptic flow, which
break the $T$--$\betaavg$ degeneracy~\cite{ALICE:2015v2}. Multistrange-hadron
measurements from the RHIC beam-energy scan~\cite{STAR:2026axo} would extend
the test to collision energies where the hadronic phase is relatively more
important.

\section{Summary}
\label{sec:summary}

We have tested whether the $\phi(1020)$ meson shares the kinetic
freeze-out of $\pi$, $K$ and $p$ in Pb--Pb collisions at
$\sqrts = 2.76$~TeV, with a common blast-wave implementation, a nested
likelihood-ratio test, and explicit propagation of the bulk fit-range and
species variations and of the $\phi$ systematic-uncertainty correlations.
In 0--5\% central collisions the $\phi$ prefers a smaller flow velocity
than the bulk at $\dchi^2 = 8.1$ ($2.1\sigma$), within a range of
$0.9$--$3.0\sigma$ under the variations considered. The tension is between
the $\phi$ and the protons and is consistent with the measured $p/\phi$
ratio. It is absent in 10--30\% central collisions. A vHLLE+SMASH hybrid
calculation analyzed in the same way predicts a smaller $\phi$ flow velocity
($-0.034$), and the data agree in direction at about two-thirds of the
predicted size. In the model, the effect combines early decoupling of
primordial $\phi$ with soft regeneration and reconstruction losses. A box
calculation shows that elastic rescattering alone cannot keep the $\phi$ in
kinetic equilibrium, while regeneration and absorption replace them with
$\phi$ that follow the medium. The same test applied to the $\Xi$ and
$\Omega$ in 0--10\% central collisions, with a design fixed in advance,
separates both from the $\pi/K/p$ freeze-out at high significance. As for
the $\phi$, the protons anchor the separation. A decays-only calculation shows
that the test is valid for these species, and the full hybrid calculation
reproduces the size of the separation in 0--10\% and 20--40\%. For the $\Omega$, which SMASH does
not regenerate, it arises from a weak coupling to the late flow.
Multistrange baryons thus show species-dependent kinetic decoupling,
while the $\phi$ lies closer to the bulk than either they or the
model: in 0--10\% and 20--40\% central collisions the data show about half
of the predicted $\phi$ separation.

\begin{acknowledgments}
We thank the ALICE Collaboration for making the data available through
HEPData.
\end{acknowledgments}

\paragraph*{Data availability.}
All experimental data used in this analysis are publicly
available~\cite{ALICE:2013mez,ALICE:2014jbq_phi,ALICE:2014jbq}. The
analysis code is available from the corresponding author upon reasonable
request.

\appendix
\raggedbottom

\section{Implementation and validation}
\label{app:validation}

\emph{Bulk reproduction.} With both charges and the ALICE default ranges,
our bulk fit reproduces the published central values and their first
(fit) uncertainties within the quoted precision (Sec.~\ref{sec:bulk}). Fitting
only the positive charges gives $(95.5~\text{MeV}, 0.652, 0.70)$, so the
choice has no practical effect. The bulk invariant yields are modeled as the
bin average of $d^2N/(d\pT\,dy)$ divided by $2\pi$ times the bin-center
$\pT$, the convention of the published tables.

\emph{Spectrum conventions.} The identified-hadron and $\phi$ spectra are
published in different conventions: invariant yields for $\pi$, $K$ and $p$,
and $d^2N/(d\pT\,dy)$ for the $\phi$, so Eq.~(\ref{eq:bgbw}) is multiplied
by $2\pi\pT$ for the $\phi$. Table~\ref{tab:jacobian} shows how the $\phi$
fit depends on this factor and on the treatment of the wide $\phi$ bins.
Without the factor, the model is softer by one power of $\pT$ and the fit
compensates with a larger flow velocity: the $\phi$ minimum moves above the
bulk flow velocity, and $\dchi^2_{\rm fp}$ rises from 8.5 to 21.5.
Evaluating the model at bin centers instead of averaging over the bins
lowers $\dchi^2_{\rm fp}$ to 5.8. The $\Xi$ and $\Omega$ spectra, published
as $d^2N/(d\pT\,dy)$, are treated in the same way as the $\phi$
(Sec.~\ref{sec:multistrange}).

\emph{Numerical checks.} Doubling the number of integration points changes
the fits and $\dchi^2$ by less than $10^{-4}$. Each minimum is found by a
grid scan followed by a local minimization, and the bulk-fit uncertainties
follow from the curvature of $\chi^2$ at the minimum.

\onecolumngrid
\setlength{\intextsep}{6pt plus 2pt minus 2pt}
\begin{table}[H]
\hsize=\textwidth\columnwidth=\textwidth\linewidth=\textwidth
\caption{0--5\% $\phi$ fit versus the $\pT$ Jacobian and bin averaging
($n = 0.712$, quadrature covariance); the last row is the default.\label{tab:jacobian}}
\begin{ruledtabular}
\begin{tabular}{l l c c}
Jacobian & Evaluation & $(T_\phi, \betaavg_\phi)$ & $\dchi^2_{\rm fp}$ \Tstrut\Bstrut\\
\hline
omitted  & bin center   & (92, 0.678)  & 21.5 \Tstrut\\
included & bin center   & (149, 0.591) & 5.8 \\
included & bin average  & (132, 0.602) & 8.5 \Bstrut\\
\end{tabular}
\end{ruledtabular}
\end{table}

\section{Full tables}
\label{app:tables}

Tables~\ref{tab:app_cent}--\ref{tab:app_regen} give the numbers behind
Secs.~\ref{sec:variations}--\ref{sec:cent}.

\begin{table}[H]
\caption{Per-centrality $\phi$ fit at the bulk $n$: quadrature best fit and
$\chi^2_{\rm min}$ (six points, three degrees of freedom), and fixed-point
$\dchi^2$ for the covariance models of Sec.~\ref{sec:cov} ($L$ in GeV/$c$).
Bulk parameters as in Table~\ref{tab:cent}.\label{tab:app_cent}}
\begin{ruledtabular}
\begin{tabular}{l c c c c c c c c}
 & \multicolumn{3}{c}{Quadrature fit} & \multicolumn{5}{c}{$\dchi^2_{\rm fp}$} \\
Centrality & $T_\phi$ (MeV) & $\betaavg_\phi$ & $\chi^2_{\rm min}$ & quad. & smooth(0.5) & smooth(1.0) & smooth(2.0) & tilt / corr. \Tstrut\Bstrut\\
\hline
0--5\%    & 132 & 0.602 & 1.09 & 8.5 & 6.0 & 6.0 & 8.5 & 5.3 / 13.4 \Tstrut\\
5--10\%   & 187 & 0.551 & 0.53 & 4.5 & 3.3 & 4.3 & 6.2 & 4.9 / 9.0 \\
10--20\%  & 134 & 0.603 & 0.78 & 2.1 & 1.5 & 1.6 & 2.7 & 2.0 / 4.1 \\
20--30\%  & 164 & 0.561 & 0.31 & 2.2 & 1.7 & 2.0 & 2.8 & 2.4 / 3.6 \\
30--40\%  & 339 & 0.330 & 0.81 & 4.9 & 4.5 & 6.5 & 9.2 & 10.1 / 8.9 \\
40--50\%  & 340 & 0.305 & 2.03 & 2.9 & 2.8 & 3.2 & 3.9 & 4.0 / 4.0 \\
50--60\%  & 190 & 0.484 & 0.67 & 3.4 & 3.3 & 3.3 & 3.9 & 2.2 / 6.0 \\
60--70\%  & 269 & 0.377 & 0.45 & 7.8 & 7.4 & 7.2 & 8.5 & 5.5 / 11.4 \\
70--80\%  & 225 & 0.402 & 2.25 & 11.2 & 10.0 & 9.1 & 10.6 & 6.5 / 14.3 \Bstrut\\
\end{tabular}
\end{ruledtabular}
\end{table}

\begin{table}[H]
\begin{minipage}[t]{0.485\textwidth}
\caption{$\phi$ fit-window variations, 0--5\% (bin centers; contiguous
windows with at least five of the eight points). The asterisk marks the
default.\label{tab:app_win}}
\begin{ruledtabular}
\begin{tabular}{c c c c c c}
Window (GeV/$c$) & $N$ & $(T_\phi, \betaavg_\phi)$ & $\chi^2_{\rm min}$ & quad. & corr. \Tstrut\Bstrut\\
\hline
0.65--2.25 & 5 & (39, 0.655) & 0.03 & 6.8 & 10.8 \Tstrut\\
0.65--2.75 & 6 & (111, 0.619) & 1.08 & 7.9 & 12.1 \\
0.65--3.50 & 7 & (125, 0.610) & 1.17 & 10.6 & 15.6 \\
0.65--4.50 & 8 & (175, 0.567) & 4.92 & 10.3 & 13.2 \\
0.90--2.75 & 5 & (118, 0.612) & 1.04 & 6.4 & 10.4 \\
0.90--3.50$^*$ & 6 & (132, 0.602) & 1.09 & 8.5 & 13.4 \\
0.90--4.50 & 7 & (204, 0.530) & 4.37 & 8.9 & 11.9 \\
1.25--3.50 & 5 & (172, 0.553) & 0.74 & 5.8 & 9.5 \\
1.25--4.50 & 6 & (349, 0.287) & 3.07 & 7.6 & 10.2 \\
1.75--4.50 & 5 & (343, 0.292) & 3.03 & 4.6 & 6.0 \Bstrut\\
\end{tabular}
\end{ruledtabular}
\end{minipage}\hfill
\begin{minipage}[t]{0.485\textwidth}
\caption{Leave-one-out test, 0--5\%: the $\phi$ fit (quadrature) and
fixed-point $\dchi^2$ with each of the six points dropped in turn.\label{tab:app_loo}}
\begin{ruledtabular}
\begin{tabular}{c c c c}
Dropped $\pT$ (GeV/$c$) & $(T_\phi, \betaavg_\phi)$ & quad. & corr. \Tstrut\Bstrut\\
\hline
0.90 & (172, 0.553) & 5.8 & 9.5 \Tstrut\\
1.25 & (133, 0.603) & 5.9 & 10.3 \\
1.75 & (151, 0.584) & 7.8 & 13.0 \\
2.25 & (104, 0.629) & 7.7 & 12.3 \\
2.75 & (134, 0.600) & 7.8 & 12.2 \\
3.50 & (118, 0.612) & 6.4 & 10.4 \Bstrut\\
\end{tabular}
\end{ruledtabular}

\vspace{0.3\baselineskip}
\caption{Regeneration scan, 0--5\% (Sec.~\ref{sec:regen}): fixed-point
$\dchi^2$ versus the regenerated fraction $f$, for a regenerated shape at
the bulk point and at the bulk flow with $T - 20$~MeV.\label{tab:app_regen}}
\begin{ruledtabular}
\begin{tabular}{c c c c c}
 & \multicolumn{2}{c}{Bulk point} & \multicolumn{2}{c}{Bulk flow, $T - 20$~MeV} \\
$f$ & quad. & corr. & quad. & corr. \Tstrut\Bstrut\\
\hline
0.00 & 8.5 & 13.4 & 8.5 & 13.4 \Tstrut\\
0.08 & 8.5 & 13.3 & 7.4 & 11.9 \\
0.15 & 8.4 & 13.3 & 6.6 & 10.7 \\
0.22 & 8.4 & 13.3 & 5.8 & 9.5 \\
0.29 & 8.4 & 13.2 & 5.0 & 8.4 \Bstrut\\
\end{tabular}
\end{ruledtabular}
\end{minipage}
\end{table}

\twocolumngrid

\bibliography{phi_prc}

\end{document}